\documentclass[12pt,a4paper,final]{iopart}

\usepackage{iopams}  
\usepackage{graphicx}
\usepackage[breaklinks=true,colorlinks=true,linkcolor=blue,urlcolor=blue,citecolor=blue]{hyperref}

\expandafter\let\csname equation*\endcsname\relax
\expandafter\let\csname endequation*\endcsname\relax
\usepackage{amsmath}
\usepackage{tikz}
\usepackage{color}
\usetikzlibrary{arrows, automata}

\begin{document}

\title[Comment on "Improvements for drift-diffusion plasma fluid models with explicit time integration"]{Comment on "Improvements for drift-diffusion plasma fluid models with explicit time integration"}

\author{Jiayong Zou}
\address{Southwest Electric Power Design Institute Co., Ltd., China Electric Power Engineering Consulting Group, Chengdu 610021, China}
\eads{\mailto{zoujiayong\_chn@126.com}}

\begin{abstract}
Recently, J. Teunissen reported a fully explicit method, namely the current-limit approach, which claimed to overcome the dielectric relaxation time restriction for the drift-diffusion plasma fluid model. In this comment, we point out that the current-limit approach is not mathematically consistent, and discuss about the possible reason why the inconsistency was not visibly noticed.
\end{abstract}

\vspace{2pc}
\noindent{\it Keywords}: fluid model, plasma, consistency, current-limit approach
%\submitto{\PSST}
%\maketitle
\newline

{\color{red} a short note highlighted in red is added on 31 July, 2020.

Recently, this note was accepted by Plasma Sources Sci. Technol., together with the response by J. Teunissen. In the response, it was claimed that the consistency should be discussed when both $\triangle t$ and $\triangle x$ approach 0. When these two parameters approach 0, the current-limiting approach turns off, thus the method is consistent. We in fact disagree with this argument. 

Considering a simple problem,
\begin{eqnarray}
\frac{\partial n}{\partial t}+\frac{\partial n}{\partial x}=0, x\in [0,2\pi]
\end{eqnarray}
with $n(x,0) = 1000000\sin(x)$ and $n(0,t)=n(2\pi,t)$. Clearly, the exact solution is $n(x,t)=1000000\sin(x-t)$. 

Following the current-limiting method, we solve it using the following scheme:
\begin{eqnarray}
\frac{\partial n}{\partial t}+\frac{\widehat{f_{i+1/2}}-\widehat{f_{i-1/2}}}{\Delta x}=0, \\
\widehat{f_{i+1/2}}=\begin{cases} \frac{1}{\Delta t} & f_{i+1/2}>\frac{1}{\Delta t}\\
f_{i+1/2} & f_{i+1/2}\leq \frac{1}{\Delta t}
\end{cases}
\end{eqnarray}
where $f_{1+1/2}$ is an upwind flux.
The above scheme should work if the current-limiting approach was correct. Does anyone think the above scheme will give correct results?

We are not willing to submit a comment to the response which is a waste of reviewers' time, so we leave an additional note in this comment.
}

Recently, J. Teunissen reported a fully explicit method which claimed to overcome the dielectric relaxation time restriction for the drift-diffusion plasma fluid model \cite{teunissen}. The dielectric relaxation time restriction results from the coupling between the Poisson equation and the charge carrier transport equations. This time restriction can be removed by semi-implicit schemes \cite{apl, lin, villa1, villa2} at the price of solving a variable-coefficient elliptic equation which is generally more expensive than solving a constant-coefficient Poisson's equation. Therefore, it would be valuable if one can use a fully explicit scheme to overcome the dielectric relaxation time restriction.

The main concern of this comment is that the fully explicit method reported by J. Teunissen, namely the current-limit approach, is not mathematically consistent. 

Quoted from \cite{arnold},"Consistency of a discretization refers to a quantitative measure of the extent to which the exact solution satisfies the discrete problem. Stability of a discretization refers to a quantitative measure of the well-posedness of the discrete problem. A fundamental result in numerical analysis is that the error of a discretization may be bounded in terms of its consistency and stability."
A numerical scheme is consistent if its discrete operator converges towards the continuous operator of the PDE when $\Delta x\rightarrow 0$ and $\Delta t \rightarrow 0$, namely, the truncation error should vanish.

Without loss of generality, we omit the diffusion term in the fluid model. The transport equations in the fluid model may be written as
\begin{equation}
\frac{\partial n}{\partial t}+\frac{\partial f}{\partial x}=s(n),\label{eq1}
\end{equation}
and the semi-discretized form of Eq. (\ref{eq1}) is, 
\begin{equation}
\frac{dn_i}{dt}+\frac{f_{i+1/2}-f_{i-1/2}}{\Delta x}=s(n_i). \label{eq2}
\end{equation}
Then, Eq. (\ref{eq2}) is consistent if it always converges to Eq. (\ref{eq1}) as $\Delta x\rightarrow 0$, namely
\begin{equation}
\frac{dn_i}{dt}+\frac{f_{i+1/2}-f_{i-1/2}}{\Delta x}=s(n_i)+O(\Delta x^p), ~~\text{with~~} p>0.
\end{equation}

According to the current-limit approach, if the flux $f_{i+\frac{1}{2}}> f_{\text{max}}$ where $f_\text{max}=\frac{\varepsilon_0 E}{e\Delta t}$ (see Eqs. (17) and (18) of \cite{teunissen}), the flux $f_{i+\frac{1}{2}}$ is limited to be $\widehat {f_{i+\frac{1}{2}}}= f_\text{max}$, and Eq. (\ref{eq2}) reads
\begin{equation}
\frac{dn_i}{dt}+ \frac{\widehat {f_{i+1/2}}- f_{i-1/2}}{\Delta x}=s(n_i). \label{eq4}
\end{equation}

Comparing Eqs. (\ref{eq4}) and (\ref{eq2}), when the current-limit approach is turned on, another problem is solved
\begin{equation}
\frac{dn_i}{dt}+\frac{f_{i+1/2}- f_{i-1/2}}{\Delta x}=s(n_i) + \frac{ f_{i+1/2}-f_\text{max}}{\Delta x},
\end{equation}
namely, an additional term $\frac{ f_{i+1/2}-f_\text{max}}{\Delta x}$ is added to the source term. Moreover, when both $f_{i+1/2}$ and $f_{i-1/2}$ are greater than $f_\text{max}$, both of them will be limited, therefore, Eq. (\ref{eq2}) becomes
\begin{equation}
\frac{dn_i}{dt}+ \frac{{f_\text{max}}- f_\text{max}}{\Delta x}=s(n_i)\Rightarrow \frac{dn_i}{dt}=s(n_i),
\end{equation}
which is obviously different from the initial problem (Eq. (\ref{eq1})), namely, the convection term is directly omitted in the computation. In either case, the current-limit approach does not converge to Eq. (\ref{eq1}) which implies that the scheme is not consistent.

In \cite{teunissen}, a convergence order in time was numerically observed (Fig. 3 and Fig. 5 in \cite{teunissen}). We present the following example to show this is possible in a numerical experiment even for an inconsistent approach:
\begin{equation}
\frac{dy}{dt}=y, ~~~y(0) = 1. \label{eq6}
\end{equation}
We solve Eq. (\ref{eq6}) with the following scheme until $t=1$,
\begin{equation}
\frac{y^{n+1}-y^n}{\Delta t}=1.001\times y^n,~~~y(0)=1. \label{eq7}
\end{equation}
Results in Tab. 1 show a numerically observed first order convergence. However, Eq. (\ref{eq7}) is not consistent with Eq. (\ref{eq6}), but is consistent with another different problem similar to Eq. (\ref{eq6}):
\begin{equation}
\frac{dy}{dt}=y+0.001y, ~~~y(0) = 1. 
\end{equation}

\Table{\label{t1} Numerical convergence rate for Eq. (\ref{eq6}) solved with Eq. (\ref{eq7})}
\br
\ns
$\Delta t$ & error ($|\exp(1)-y^h(1)|$)& numerical convergence rate \\
\mr
0.04 & $0.04988$ & $ $ \\
0.02 & $0.02405$ & $1.05$ \\
0.01 & $0.01079$ & $1.16$ \\
0.005 &0.004065 & 1.41 \\
\br
\end{tabular}
\end{indented}
\end{table}

Now we discuss on the possible reason for why there were no visible differences between the current-limit approach and the explicit scheme shown in \cite{teunissen}. In the fluid model, $f=nv$ and $s(n)=\alpha n|v| = \alpha |f|$, with $\alpha$ not small, and typically $\alpha\gg 1$. When both $f_{i+1/2}$ and $f_{i-1/2}$ are limited, $f_{i-1/2}-f_{i+1/2}$ may cancel to a large degree; when $f_{i-1/2}< f_\text{max}<f_{i+1/2}$ (i.e., only $f_{i+1/2}$ is limited), because $f_{i+1/2}$ and $f_{i-1/2}$ are the fluxes of a same cell, they are generally close to each other, therefore, $f_\text{max}$ is close to $f_{i+1/2}$. In either case, the additional term may be much smaller than $s(n_i)$. Therefore, the inconsistency may not be visibly noticed. This coincides with the observation in the example of Eq. (\ref{eq6}) and Eq. (\ref{eq7}).

Finally, we wish to emphasize that in this comment we focus on the mathematical characteristics of the current-limit approach itself, not on a possible way to get a visually similar result. We feel that a mathematically correct scheme is preferred for reliable simulations.

\section*{References}
%\bibliography{ssnalbib}{}
%\bibliographystyle{plain}

\end{document}